\documentclass[aps,eprint,amsmath,amssymb,floatfix,showpacs,twocolumn]{revtex4-1}
\usepackage{graphicx}
\usepackage{dcolumn}
\usepackage{bm}
\begin{document}
\title{Effect of surface roughness on the Goos-H\"{a}nchen shift}
\author{Z. Tahmasebi}
\author{M. Amiri}
\email{m\_amiri@basu.ac.ir}
\affiliation{Department of Physics, Bu-Ali Sina University,\\Hamedan 65175-4161, Iran}
\begin{abstract}
By considering an optically denser medium with a flat surface, but with natural roughness instead of abstract geometrical boundary which leads to mathematical discontinuity on the boundary of two adjacent stratified media, we have thus established the importance of considering physical surfaces; and thus we studied the Goos-H\"{a}nchen (GH) effect by ray-optics description to shed light on parts of this effect which have remained ambiguous. We replaced the very thin region of surface roughness by a continuous inhomogeneous intermediary medium. Applying Fermat's principle for the incident light ray, few fundamental questions about GH shift are more convincingly addressed which are in excellent agreement, even with the most details of the experimental results.
\end{abstract}
\maketitle
\section{INTRODUCTION}
The GH effect \cite{GooHan47} has been the subject of many investigations since its discovery \cite{Art48,Ren64,Agu68,Lot68,HorTam71,BeaImb73,McGCar77,RhoCar77,RisVig92,LaiXu00,JiaWan08,SchWan08,WanZhu09} and has been extended to many areas of physics \cite{CheLi10,WuCha11,JiaWen13,HaaWel10,AssXia11}, and to a wide range of materials \cite{MerWoe07,HeHe06,FelSma03,LaiCha02,WilGil82,Ber02,ShaKiv03,WanZhu05}. Our effort in this work aims to add a bit of ray-optics description to what have been obtained from wave theory about the effect in one respect in which we think it is incomplete. Based on the theory of Maxwell Garnett \cite{MaxGar04}, a rough surface separating two dielectric bulk media can be replaced by three media composed of two different homogeneous media with well-defined refraction indices  and an equivalent intermediary inhomogeneous medium with plane parallel boundaries. In general, two intermediary-medium models can be used: (i) a homogeneous-intermediary model and (ii) an inhomogeneous-intermediary model. The first model requires an effective fixed refractive index and the second model requires variability in refractive index. Several theoretical models utilizing linear variation of the refractive index, have been carried out by a few workers \cite{KozHar78,DelDel03}; but that these models agree fairly well with experiments. In contrast to earlier works, we assumed that the separating surface of the two optically dense and less dense media is a natural surface consisting of a random distribution of roughness, the deviation of which relative to an average surface is best described using the statistical distribution. This assumption leads to a nonlinear behavior of the refractive index by transition from one medium to the other and consequently the obtained behavior of the GH shift will approach more and more closely even with the most details of the experimental results.\\
\section{STATISTICAL ANALYSIS OF ROUGH SURFACE AND EQUIVALENT REFRACTIVE INDEX}
Roughness as an inseparable feature of almost all known surfaces is a macroscopic deviation from flatness. However, almost always, this deviation is random except for some regular features which have been deliberately introduced. The statistical parameters to characterize the surface topography, regardless of their sources and irrespective of the absolute scale of the size involved, look very similar \cite{Whi02,GadSol02}. From a statistical point of view, surface roughness profiles in one dimension resemble electrical recordings of white noise. Bearing area curve (BAC) is a statistical method for the analysis of surface topography \cite{StaBat14}. From Fig. 1(a), BAC could be found by considering the fraction of the random surface profile intersected by an infinitesimally thin plane with thickness $\Delta z$ positioned at height $z$ and parallel to the $x, y$-plane.
\begin{figure}[htb]
\includegraphics[width=0.25\textwidth]{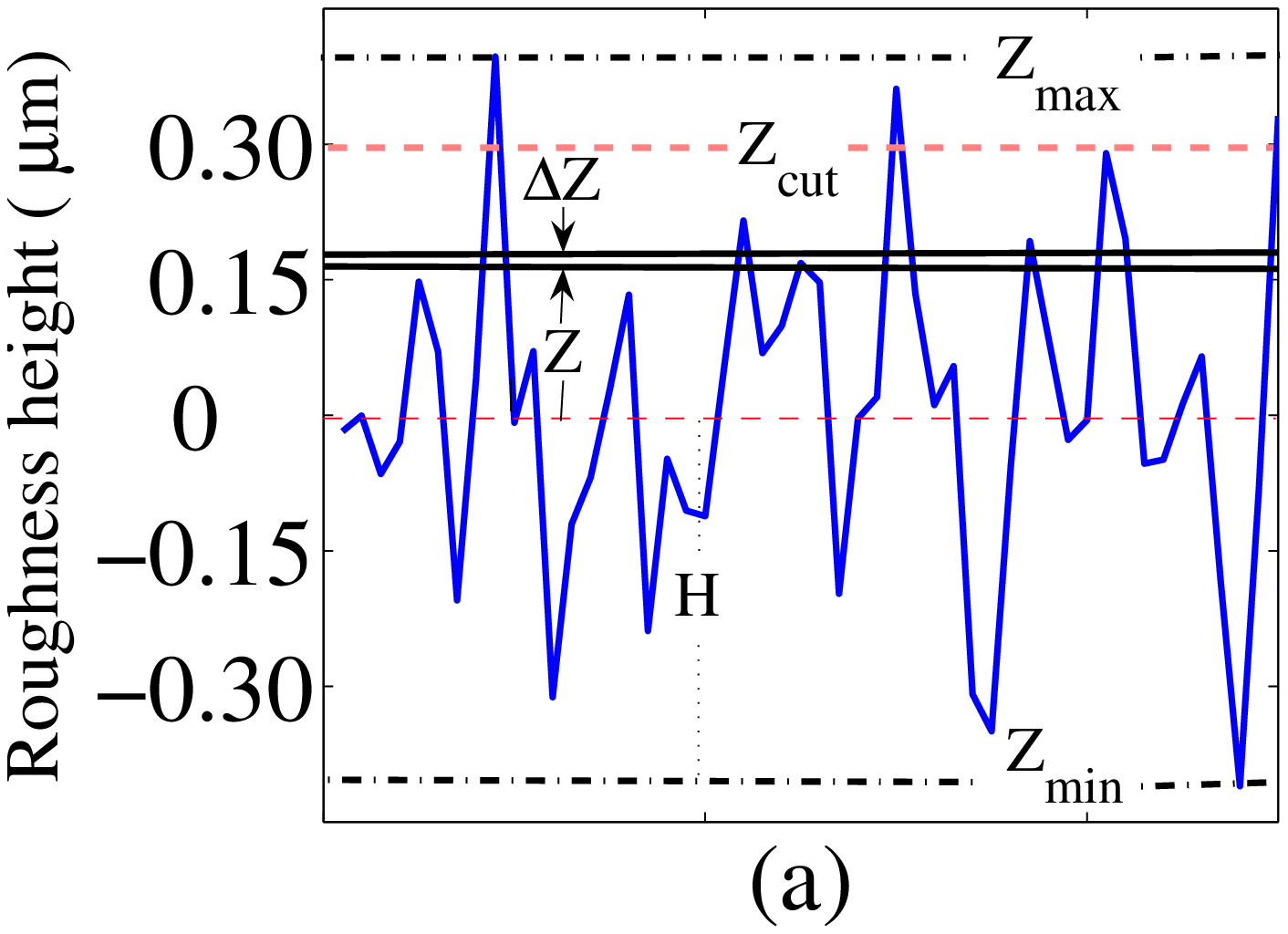}\\
\includegraphics[width=0.25\textwidth]{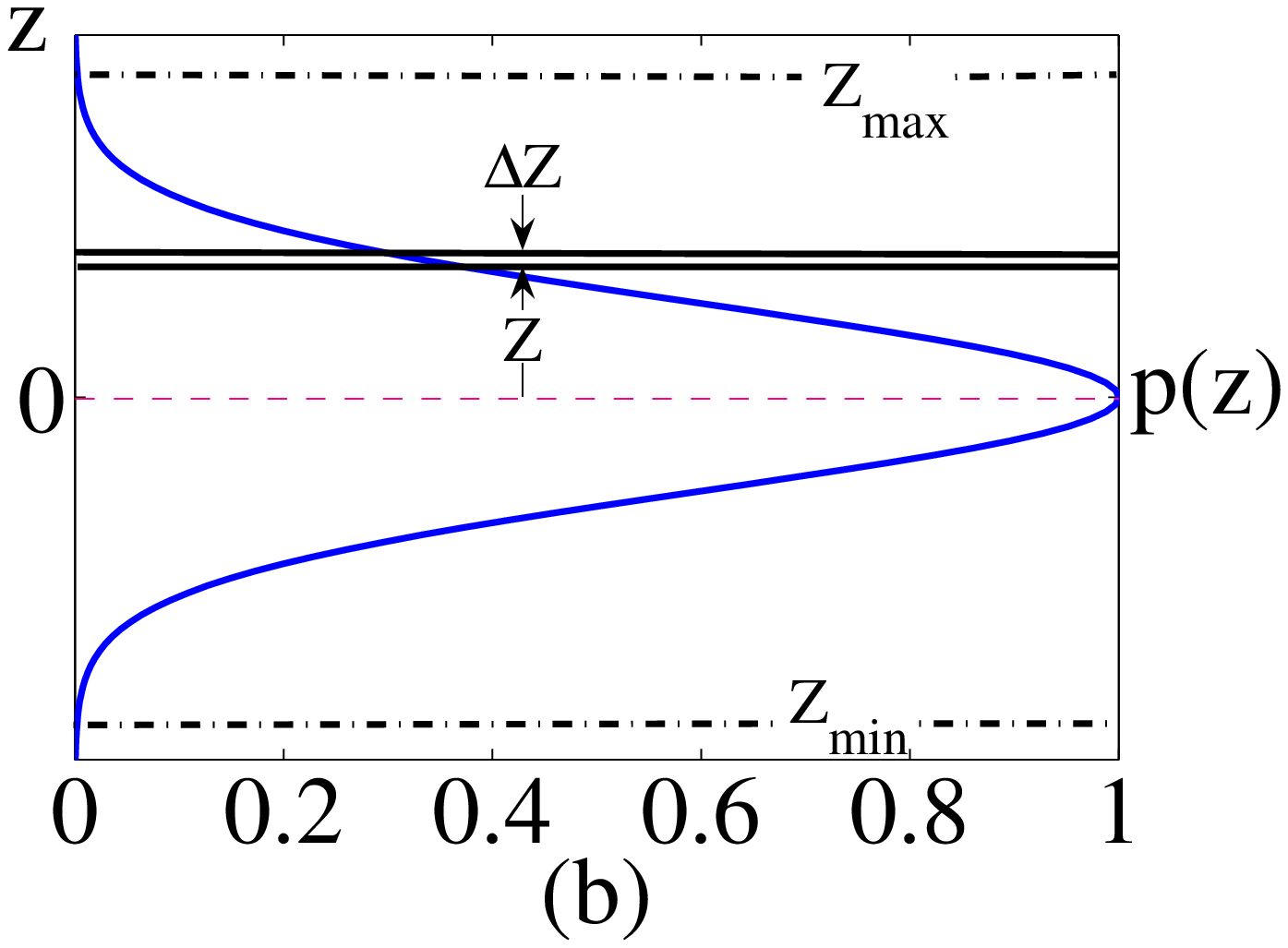}\includegraphics[width=0.25\textwidth]{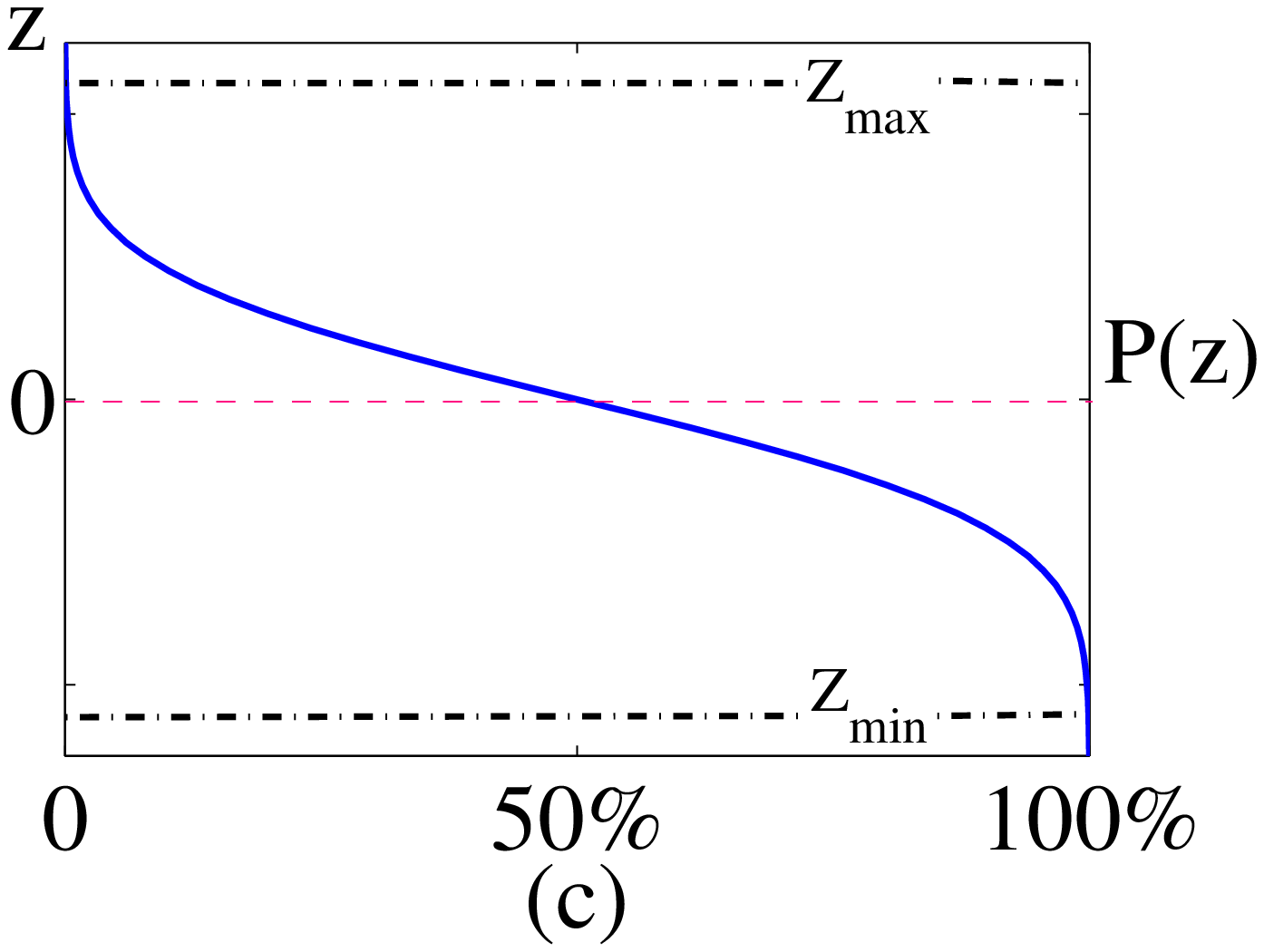}
\caption{Graphical representation of the determination of a BAC; (a) Hypothetical cross section of a flat surface with its roughnesses, $z$ is the distance perpendicular to the plane of the surface, $\Delta z$ is the interval between two heights, $H$ is the value of average height, (b) $p(z)$ is the probability density function of roughness, and (c) $P(z)$ is the integral of the $p(z)$ \cite{StaBat14}.}
\end{figure}
The procedure is repeated through a number of slices from a lowermost plane at $Z_{min}$ to an uppermost one at $Z_{max}$. To evaluate the equivalent refractive index for roughness region, we use the following process: First, we must determine the probability density function of roughness, $p(z)$. Next, we obtain the BAC by integrating the height probability density function, $P(z)$, which corresponds to the presence of matter in the height $z$ and consequently the refractive index at that height. For Gaussian height distributions, $P(z)$ is the cumulative probability function of classical statistics. Suppose that the roughness is a function of two variables $x$ and $y$. If the plane of the chief ray of the incident beam (or the center of gravity of the incident beam), is considered to be parallel to $x, z$-plane, we obtain a graph from its intersection with the rough surface in that plane which gives a continuous random function as $z(x)$. Evidently, at an arbitrary plane, there will be numerous different roughness profiles. For each $z(x)$, we can assign a continuous probability distribution function, $p(z)$. As the rough surface is scanned along the $x$ axis at height $z$, $p(z)$ shows recording the value of height $z$ during the scan. If data approach a wide set of random phenomena, probability density function approaches Gaussian distribution \cite{Ste90}. Therefore, despite each $x, z$-plane consists of different profiles, we can legitimately assume that each of these profiles is Gaussian. Surely, the selection of any non-analytical distribution, even matched with reality, causes analytical discussion to get into trouble. A normal statistical distribution in the region of roughness with random variable $z$, mean $\mu$ and variance $\sigma ^2$ with probability density function $p(z)$ is shown as below:
\begin{equation}
p(z)= \frac{1}{\sigma\sqrt{2\pi}}exp[-\frac{(z-\mu)^2}{2\sigma^2}].
\end{equation}
Integrating the probability density function, we can write the BAC as
\begin{equation}
P(z)=\int^{+\infty}_{-\infty}p(z)dz-\int^{z}_{-\infty}p(z)dz.
\end{equation}
In the examination of a surface with random roughness, we supposed that the uppermost roughness peak was located at height $Z_{max}$ and the lowermost valley was located at height $Z_{min}$. Therefore, the average value of the height is equal to the average value of the sum of $Z_{max}$ and $Z_{min}$,  which we quite deliberately positioned at $z=0$, hence, $\mu=0$ as shown in Fig. 1(a). Then, the first integral in BAC is the contribution of all height distribution in the roughness region of the surface which is normalized to unity. But, the region of random roughness starting from $Z_{min}$ where the dense matter completely exists and ends up at $Z_{max}$ and does not exist above it. Consequently, by replacing $-\infty$ and $+\infty$ for the lower and upper limits of integration instead of $Z_{min}$ and $Z_{max}$, respectively; integration would conduce to insignificant approximate value. The second term in the BAC of matter is also the contribution of height, the value of which is less than a certain $z$.  Therefore, the result of the two integrals specifies the amount of the presence of matter along the matter to the desired height. Then, the BAC of matter becomes
\begin{eqnarray}
P(z)=1-\int^{z}_{-\infty}p(z)dz,\nonumber \\
=\frac{1}{2}[1-erf({\frac{z}{\sigma\sqrt{2}}})].
\end{eqnarray}
The bell-shaped probability density function and the BAC obtained from error function are shown in Figs. 1(b) and 1(c). In this work, the surface separating two different optical media has been considered to be a natural surface of glass which separates glass as an optically dense medium from air as a less dense one. In effect, an intermediary inhomogeneous medium is made of binary mixture that has a coexistence between glass and air. Now we can determine the equivalent refractive index of an intermediary inhomogeneous medium, but with gradual change of refractive index along $z$ axis, which varies from constant value $n_{g}$, where there is clear glass to $n_{a}=1$, where there is clear air. Furthermore, we shall assume that the refractive index is proportional to the presence of matter or BAC, then, since 
$n(Z_{min})=n_{g}$, $n(Z_{max})=1$, $Z_{max}=-Z_{min}=H$, and also error function is an odd function; we obtain the exact relation for refractive index
\begin{equation}
n(z)=\frac{(n_{g}+1)}{2}-\frac{(n_{g}-1)}{2}[\frac{erf(\frac{z}{\sigma\sqrt{2}})}{erf(\frac{H}{\sigma\sqrt{2}})}].
\end{equation}
Figure 2 shows the behavior of the variable refractive index for monochromatic light in an inhomogeneous medium.
\begin{figure}[htb]
\includegraphics[width=0.5\textwidth]{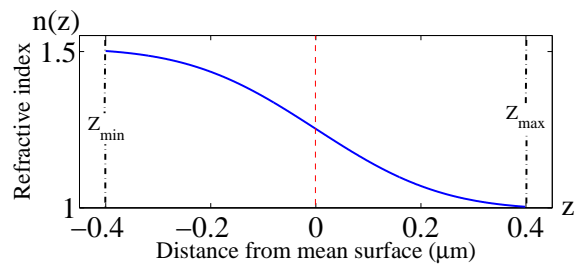}
\caption{The behavior of the refractive index in an intermediary inhomogeneous medium.}
\end{figure}
\section{RAY TRACING IN AN INTERMEDIARY MEDIUM}
Based on what was obtained from the refractive index in an inhomogeneous region as a function of height and Hamilton's principle, we try to determine the equations of motion for a light ray through an inhomogeneous medium. Although, it was known that there are some differences between the penetration depth of the beam path and that of the ray path into an inhomogeneous region \cite{KozKim76}. By means of Hamilton's variational principle, we can formulate Lagrange's equations for ray optics and obtain the equation of ray path. Here the Lagrangian can be expressed as a function of $z(x)$, $z'(x)$ and $x$. From Hamilton's principle, we derive Lagrange's equation as
\begin{equation}
\frac{d}{dx}\frac{\partial L(z, z'; x)}{\partial z'}-\frac{\partial L(z, z'; x)}{\partial z}=0, z'=\frac{dz}{dx}.
\end{equation}
To obtain the optical equivalence of $L(z, z'; x)$, according to Fermat's principle of the minimum optical path in an inhomogeneous medium\cite{BorWol99}
\begin{equation}
\int_{z_{1}}^{z_{2}}n(z)ds=\int_{z_{1}}^{z_{2}}n(z)\sqrt{1+z'^{2}}dx,
\end{equation}
we then obtain Lagrangian of the optical system as
\begin{equation}
L(z, z'; x)=n(z)\sqrt{1+z'^{2}}.
\end{equation}
On substitution $L(z, z'; x)$ into Lagrange's equation one finds
\begin{equation}
\frac{d}{dx}[n(z)\frac{z'}{\sqrt{1+z'^2}}]-\sqrt{1+z'^2}\frac{dn(z)}{dz}=0.
\end{equation}
The first term on the left side of the above equation becomes
\begin{equation}
\frac{dn(z)}{dx}\frac{z'}{\sqrt{1+z'^2}}+\frac{d(z')}{dx}\frac{n(z)}{\sqrt{1+z'^2}}+n(z)z'\frac{d}{dx}\frac{1}{\sqrt{1+z'^2}},
\end{equation}
which after some manipulation leads to
\begin{equation}
\frac{\frac{dn}{dz}}{n(z)}z'^2+\frac{d^2 z}{dx^2}\frac{1}{1+z'^2}.
\end{equation}
This result can be combined with the second term of the left side of the Eq. (8) and yielding the following equation
\begin{equation}
\frac{d^2 z}{dx^2}=\frac{\frac{dn}{dz}}{n(z)}[1+(\frac{dz}{dx})^2].
\end{equation}
$ds$ which shows the infinitesimal increments of a ray path in an inhomogeneous region is related to its components in Cartesian coordinate system, then
\begin{equation}
\frac{ds}{dx}=\sqrt{1+z'^2}=\frac{1}{\sin\theta(z)}=\frac{n(z)}{n_{g}\sin\theta_{i}}.
\end{equation}
On performing the substitution given by the preceding equation in Eq. (11), we finally obtain the ray path differential equation as
\begin{equation}
\frac{d^2 z}{dx^2}=\frac{(\frac{dn^2(z)}{dz})}{2n_{g}\sin^2\theta_{i}},
\end{equation}
which is a second-order nonlinear differential equation. To solve it, we used the Runge-Kutta method (RK4), which determines the ray path in an inhomogeneous region. To do so, we considered the incident rays from the dense homogeneous region of the glass with refractive index $n_{g}$ at different angles of oblique incidence into a medium with varying refractive index. Since it turns out from numerical considerations that the maximum value of the GH shift is-for a given standard deviation-linearly proportional to the thickness of the intermediary medium, we have depicted the ray path in an intermediary region with the thickness of $0.8$ $\mu m$ which was chosen deliberately for comparison with some other previous experimental results. Here, we have considered a Gaussian height distribution with the standard deviation which relates to $Z_{min}$ and $Z_{max}$ as: $Z_{max}-Z_{min}=4.21\sigma=0.8\mu m$. This amount of standard deviation covers about more than $96$ percent of
the roughness region. For $BK7$ glass with refractive index $n_{g}=1.52$ at wavelength $625$ $nm$, the critical angle is then $\theta_{crit}=41.14^o$ and Figs. 3(a) to 3(c), represent ray paths with different incident angles that are less than, equal to, and greater than the critical angle, respectively.
\begin{figure}[htb]
\includegraphics[width=0.25\textwidth]{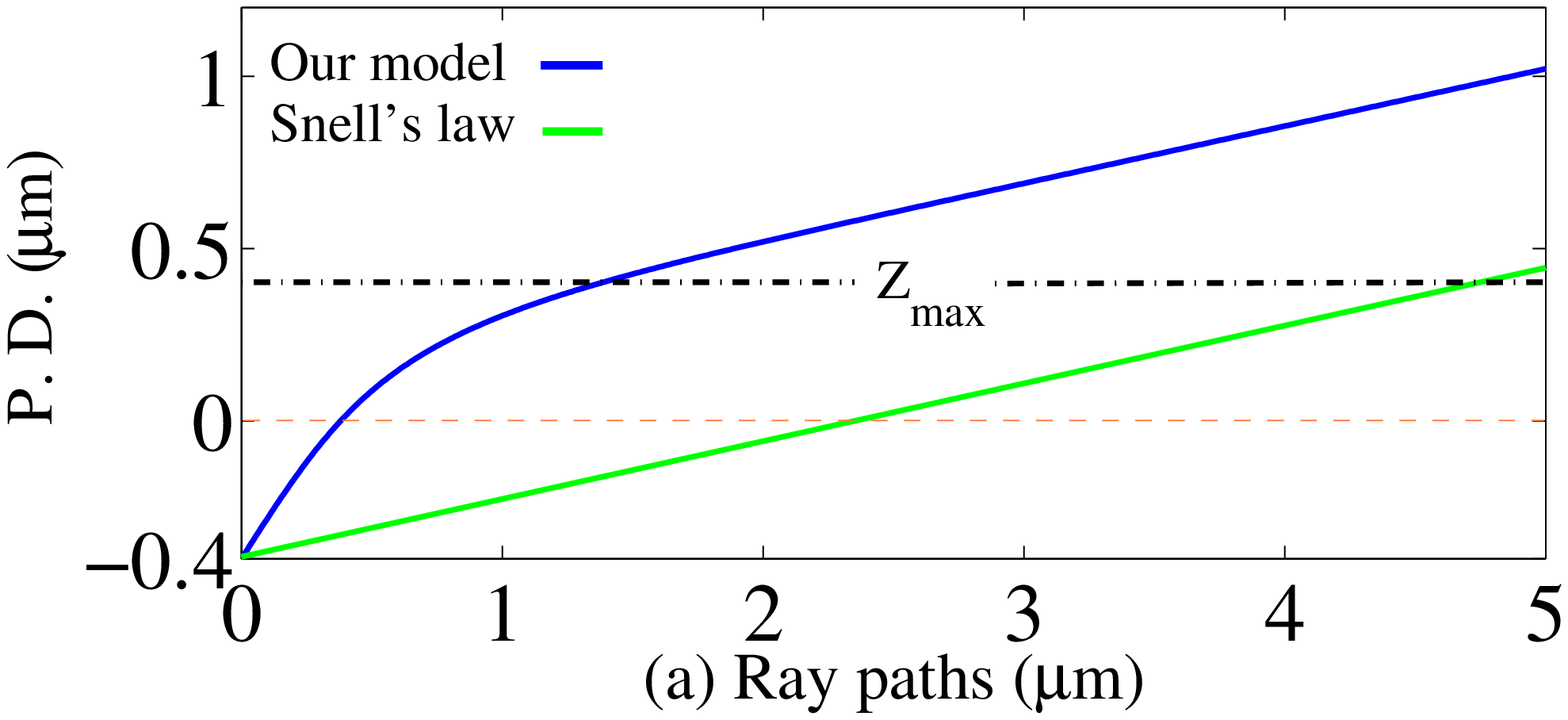}\includegraphics[width=0.25\textwidth]{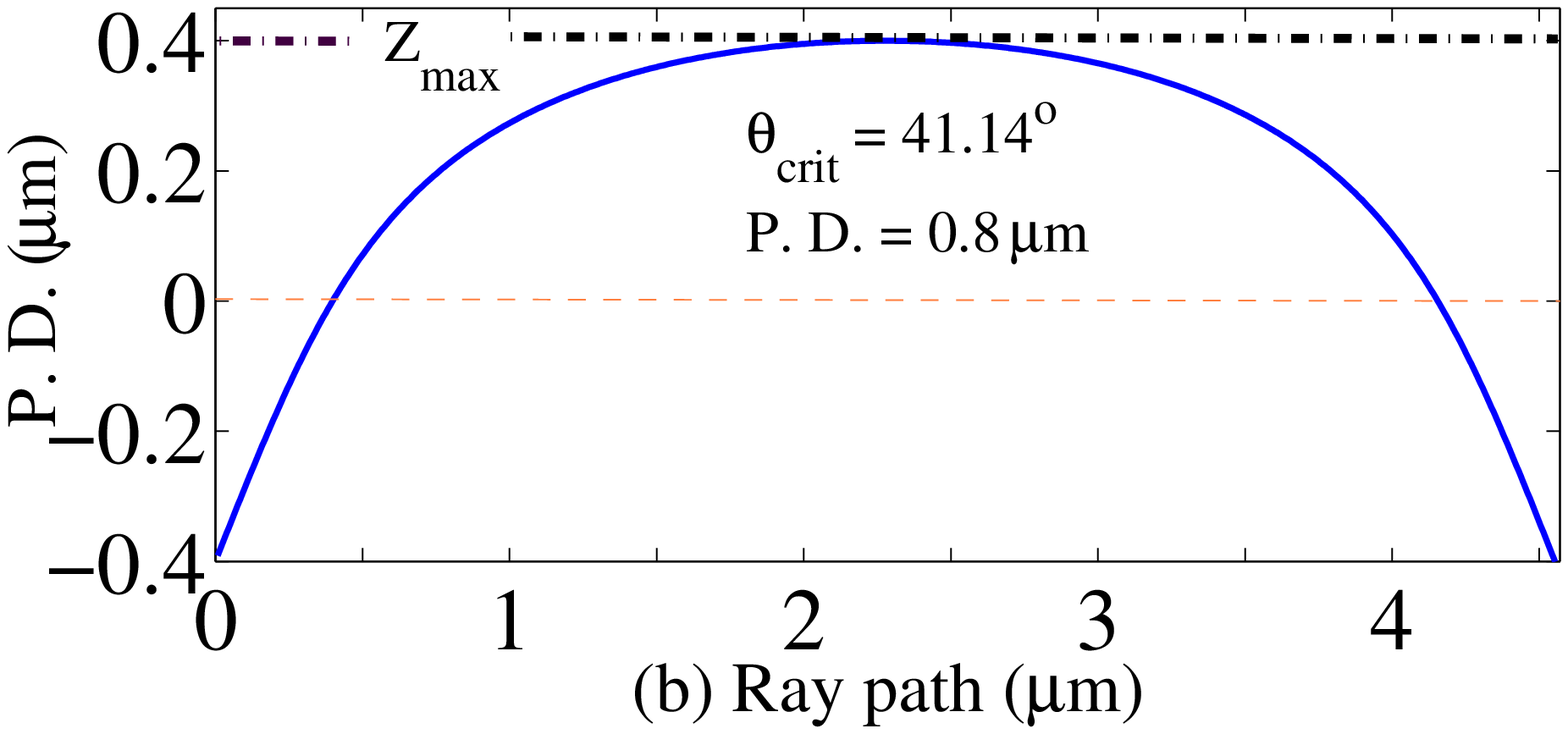}
\includegraphics[width=0.25\textwidth]{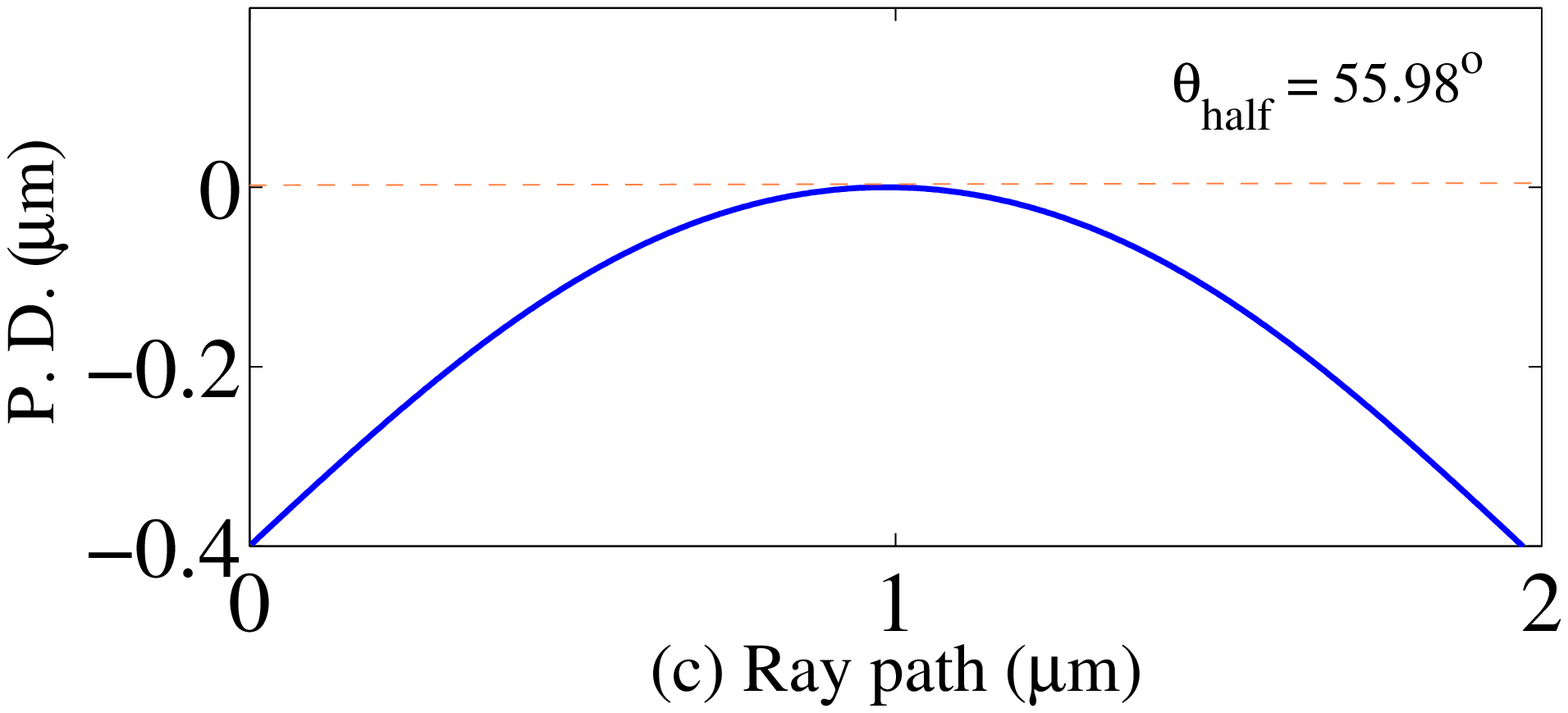}\\
\caption{Ray paths obtained numerically in an inhomogeneous region with different incident angles for $BK7$ glass with $n_{g}=1.52$ at wavelength $625nm$; (a) for $\theta_{i}<\theta_{crit}$, (b) at $\theta_{i}=\theta_{crit}=41.14^{o}$, and (c)  at $\theta_{crit}<\theta_{i}$; (P. D.) stands for penetration depth.}
\end{figure}
As can be seen in Fig. 3(a), at an incidence angle less than the critical angle; the ray passes through the inhomogeneous region along a curved path. Also, the ray path traversed by use of Snell's law in the absence of an inhomogeneous region is depicted for comparison. In Figs. 3(b) and 3(c), the ray of light arriving at the inhomogeneous region, deviates from the straight line and after travelling in a curved trajectory which does not intersect the border, bends towards the incident medium. Therefore, the emergent light ray undergoes some displacement in comparison with what is predicted by geometrical optics. What can be seen from Fig. 3(b), is that for incidence at the critical angle, a ray suffers the greatest penetration depth and GH shift. The GH shift and the penetration depth which are computed numerically as separate functions of incident angles are shown in Fig. 4(a). Also the GH shift and the penetration depth for a linear variation model of the refractive index and the penetration depth from theoretical approach \cite{Jac99} is depicted in this Figure for comparison. As can be seen from this Figure, our numerical result for the GH shift for an intermediary inhomogeneous medium with a linear variation refractive index is in perfect agreement with what has obtained in Ref. \citep{DelDel03}. But, in contrast, is very different from the one just obtained by observations, which reveals that the pattern of the roughness profile seriously matters here.
\begin{figure}[htb]
\includegraphics[width=0.5\textwidth]{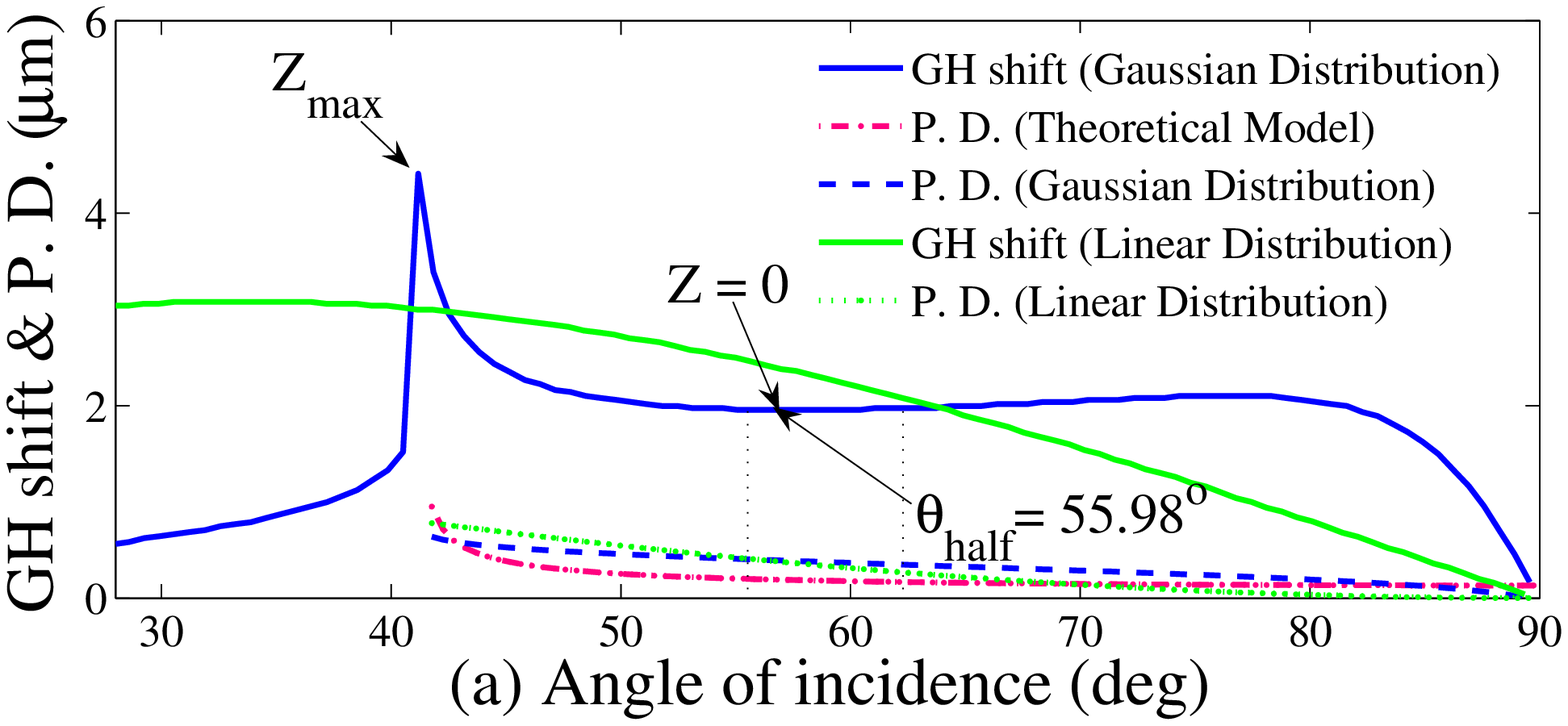}\\
\includegraphics[width=0.5\textwidth]{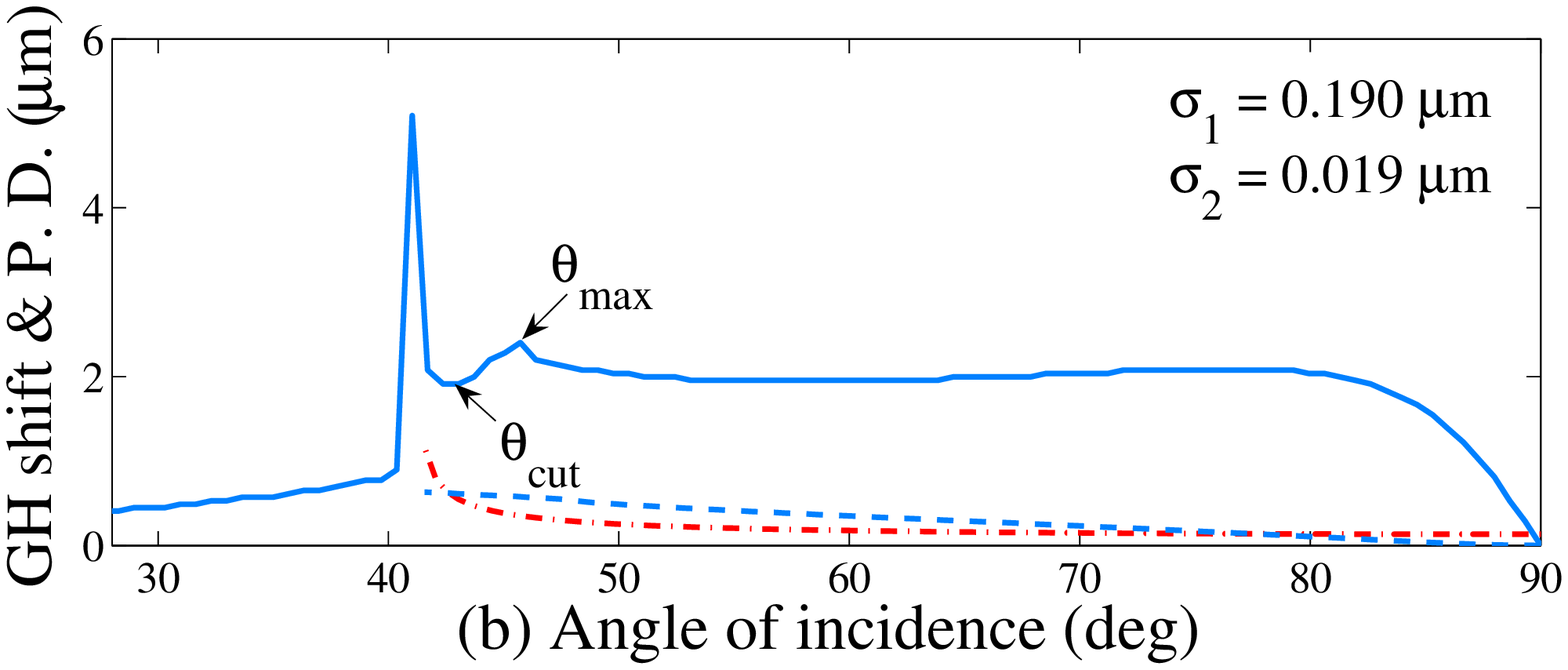}
\caption{(a) The behavior of the GH shift and the penetration depth and (b) the effect of 
the machining process of the natural surface on the GH shift and the penetration depth for 
incidence angles between $28^{o}\leq\theta_{i}\leq 90^{o}$.}
\end{figure}
With respect to Fig. 4(a), as the incidence angle grows up and nears the critical angle, it causes the GH shift picks up to a global maximum. Here, contrary to Artmann's \cite{Art48} theoretical approximation and similar to what is obtained by the Gaussian wave approximation of Horowitz and Tamir \cite{HorTam71}, the GH shift in our model has a broader shoulder and also has a limit when total reflection takes place. A further increase of $\theta_{i}$ will cause the GH shift to fall sharply and if this procedure continues numerically, for all incidence angles within the total reflection range, the GH shift at a relatively wide range of incidence angles between $52^{o}$ and $61^o$ takes to reach a minimum, thereafter the GH shift increases slowly and after reaching a wide range maximum between $74^o$ and $78^o$, falls off nearly to zero in this limit. A wide range of incidence angles instead of being a single one in the former cases, may be an essential aspect of the numerical method which has been applied. As we shall see later, these minimum and maximum are not an artifact and can be deduced from a sound physical basis. The main differences between our model with what was observed experimentally in the behavior of the GH shift at the metal-air interface \cite{SchWan08}, are the existence of the minimum and the maximum and also, as it should be, the vanishing GH shift at the intermediary-air interface at nearly grazing incidence. Unlike the GH shift, the penetration depth of a ray path will have a descending course from the greatest value at the critical angle, $P.D.\simeq2H$, to zero at the end of grazing incidence. After a closer numerical examination, however, it seems that the minimum GH shift is related to the inflection point of the refractive index profile of the inhomogeneous region around the height $z=0$. As can be seen in Fig. 2, the refractive index at $z=0$ reaches its extreme change and consequently, the change of the ray path at this region is severe and this, in turn, causes bending a ray towards the incident medium to be faster than other angles. A particularly attractive theoretical feature of this effect is the fact that at the incidence angle $\theta_{half}=55.98^o$, Fig. 3(c),  which is in between the minimum interval, the penetration depth is half the thickness of the roughness region, which may be of great practical importance for determining the statistical parameters of the surface. On the other hand, it is worth noting that slow changes in the refractive index around $Z_{min}$ besides the grazing incidence of a ray causes bending a ray towards the incident medium to be much more slowly than other angles which makes the GH shift increase and reach a wide range maximum. This case is almost similar to what happens to the incidence at the critical angle when a ray path approaches the border line at $Z_{max}$. This is because the angle of the incident ray and the normal to the intermediary-air interface before reaching the border and that of the incident ray relative to the glass-intermediary interface are the same, that is grazing. However, the major difference in the former case is that a light ray penetrates deeply through an intermediary medium with variable refractive index to the border line at $Z_{max}$ and then returns to the incident medium and in the latter one a light ray travels along the interface surface with a relatively constant refractive index and very low penetration and returns to the incident medium. After passing this maximum range, the curve in Fig. 4(a) is seen to decrease to zero. Thus, unlike the classical model which predicts that the GH shift will have an indefinite value between zero to infinity, our model predicts a zero GH shift. As can be expected and verified numerically, the distance between the fixed global maximum at $\theta_{crit}$ and the two successive minimum and maximum would change depending on the change in $\sigma$ and $H$.\\ With respect to the changes in the profile of the glass surface produced by the machining processes such as grinding and polishing, we can now achieve a more precise result by modifying our model. From the modelling point of view, we simply suppose that the crests of the roughness cutting off to a height $Z_{cut}$ after grinding and then another Gaussian distribution can be produced by the polishing process but with very small height distribution which is centered at the height $Z_{cut}$. Thus, this can easily be explained in an approximate way in terms of a combination of two normal distribution functions, but with different standard deviations $\sigma$ and $\sigma'$, which $\sigma=10\sigma'$, chosen arbitrarily (Ref. [30], p. 8). Fig. 4(b) depicts the behavior of the GH shift and the penetration depth based on the modified model and in the same range of incident angles as before. As can be seen, the modified roughness distribution causes new successive minimum and maximum to appear in addition to the former ones and immediately after $\theta_{crit}$, which have the same physical basis as before and are located at angles $\theta_{cut}$ and $\theta_{max}$, respectively. This new characteristic feature observed through direct experimental investigation \cite{SchWan08} for both TE and TM polarized light but was ignored by the authors and also was unexpected in the aforementioned models. Another point of the cardinal importance in Fig. 4(b) is that the location of the $\theta_{cut}$ corresponds to the penetration depth by a curved trajectory which will reach the height $Z_{cut}$, the inflection point of the new refractive index profile, and that is why we named it $\theta_{cut}$. Therefore, Fig. 4(b) contains two useful theoretical and practical data concerning the center of the two roughness distribution of the polished surface and that of the ground surface, that is, at $\theta_{cut}$ and $\theta_{half}$, respectively. As can be seen in Fig. 4(b), this behavior is in good agreement with direct experimental observation \cite{SchWan08} even with its fine details. Fig. 5 compares and gives another confirmation of the normalized GH shift behavior for different relative refractive indices ranging from $n_{g}= 1.02$ to $n_{g}\leq5$ at the critical  angle given by the Horowitz and Tamir model for $TE$ and $TM$ polarization with our proposed model based on the geometrical light ray approach. Consequently, as observed from Figures (4) and (5), our proposed model predicts a finite value for the GH shift both at the critical angle and at grazing incidence which is consistent with experimental observations \cite{RhoCar77,SchWan08}. 
\begin{figure}[htb]
\includegraphics[width=0.5\textwidth]{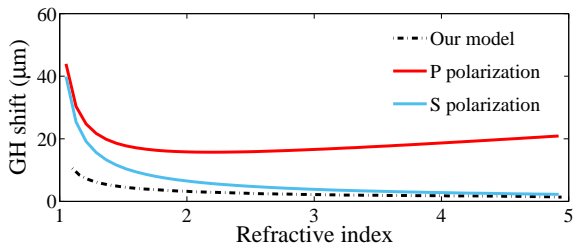}
\caption{Comparison of the normalized GH shift for incidence at $\theta_{i}=\theta_{crit}$ as a function of the relative index $n$. The solid curves are obtained from the theoretical results of a Gaussian light beam for both polarizations \cite{HorTam71} and the dashed curve obtained from our geometrical light ray approach.}
\end{figure}
Consequently, using an insufficiently accurate model we have been able to describe the detailed behavior of the GH shift in a good agreement with the experimental observation. Finally, a comparison of the results obtained for the  GH shift associated with total internal reflection is based on wave theory and our ray-optics model is worth mentioning. Many different theoretical approaches come under the heading of Maxwell's theory, have been presented so far which state that the total reflection does not take place abruptly at the boundary surface. But that due to the limited extent of the beam in its transverse direction just before the total reflection occurs, the light energy penetrates inhomogeneously and moves a short distance along the boundary within the less dense medium with the phase velocity $v_{a}/\sqrt{\varepsilon_{g}}\sin\theta_{i}$, where $v_{a}$ and $\varepsilon_{g}$, are the speed of light in air and the permittivity constant of the glass, respectively \cite{Lot70}. Consequently, this leads to a shift of the center of gravity of the totally reflected beam. On the other hand, however, what comes from our model is that an incident light ray from the denser medium enters into a very thin intermediary inhomogeneous layer with variable speed depending on the path of the light ray, and after travelling in a differentiable curved trajectory which does not enter into the air, returns to the incident medium. The speed of the light rays in the intermediary region varies with the position from the minimum value of $v_{g}$ at $Z_{min}$ to the maximum value of ${v_a}$ at $Z_{max}$. One can compare the outputs of the two approaches from Fig. 6, which depicts the situation.
\begin{figure}[htb]
\includegraphics[width=0.25\textwidth]{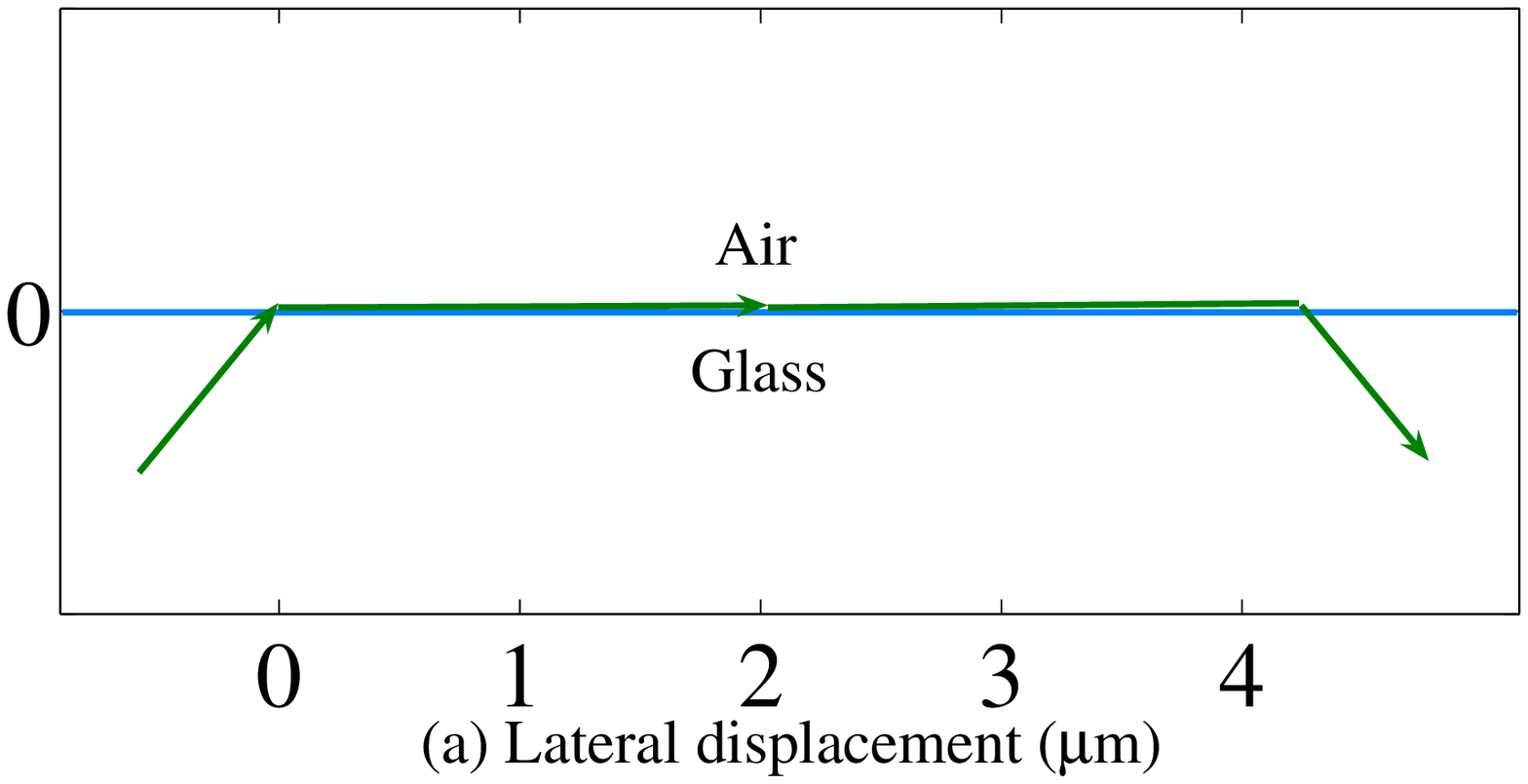}\includegraphics[width=0.25\textwidth]{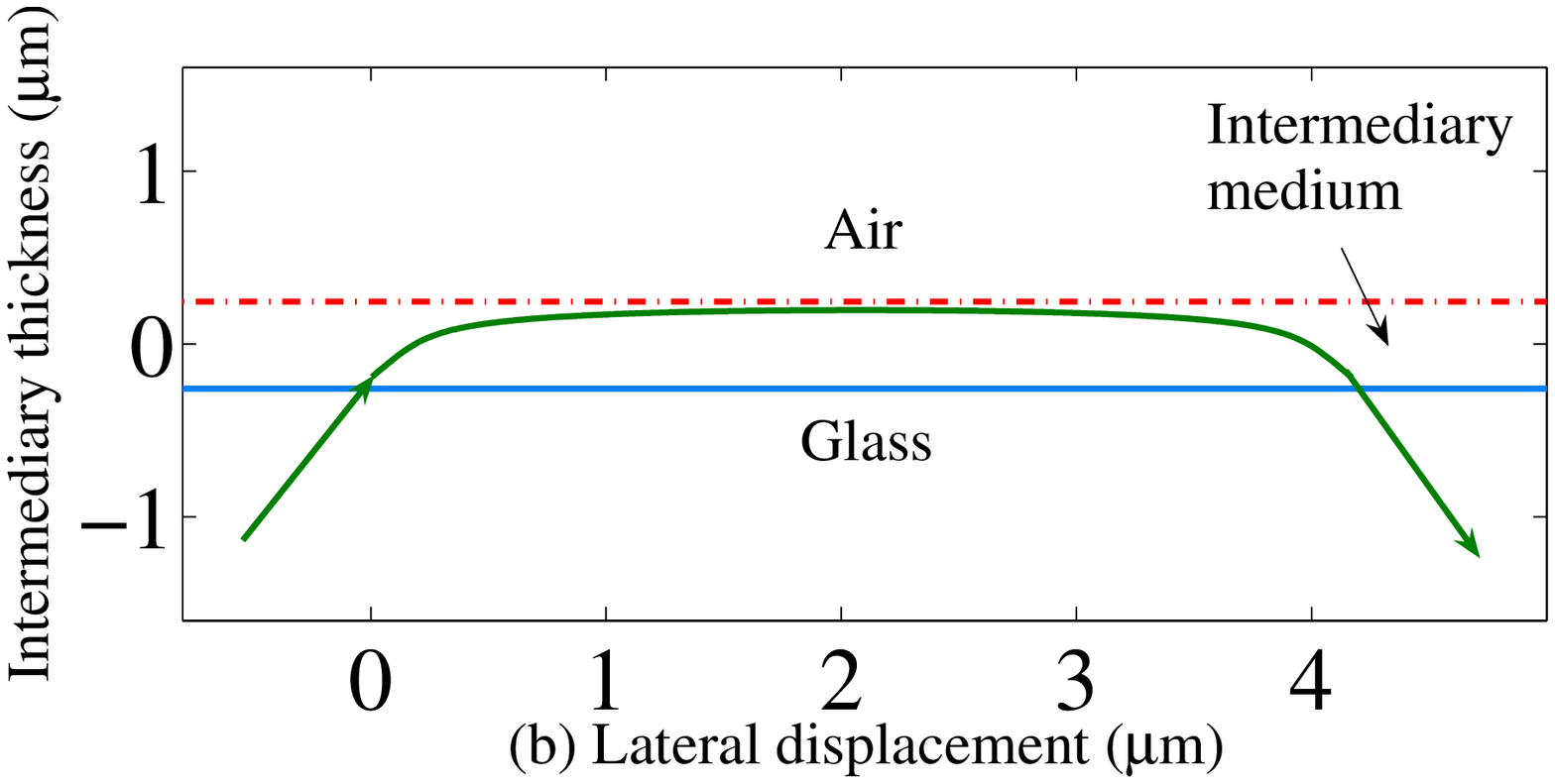}
\caption{Schematic comparison of the path of the center of gravity of the incident beam and the ray path at the same incidence angle when total internal reflection takes place; (a) from Maxwell's theory and (b) from ray-optics model, respectively.}
\end{figure} 
\section{Conclusions}
To sum up, using different approaches to study GH shift, would bring about various results all aspects of which, would certainly be impossible to cover by only one broad and comprehensive single theory or method. Utilizing ray approach, and abandoning the abstract geometrical boundary separating two different media and considering the roughness of the glass surface and replacing it with an intermediary inhomogeneous medium with graded refractive index, we could describe the GH shift in a good agreement with other models and experimental observations. At first, it is found that regardless of an intermediary medium thickness, $\theta_{crit}$ at a good approximation is independent of an inhomogeneous region thickness. However, the procedure outlined, is entirely confined to natural or artificial flat surfaces with Gaussian distribution, but if we add another Gaussian distribution with smaller standard deviation upon the cut off former one, to simulate the effects of the grinding and polishing of an optical flat surface, the obtained pattern of the GH shift versus incident angle becomes more and more similar to the observed experimental one. Using this pattern and obtaining $\theta_{cut}$ and $\theta_{half}$, it is theoretically possible to determine the position of the center of the polished and the ground distribution and also the thickness of the rough region, which can be considered as a new method to determine statistical parameters and flatness quality of a transparent medium surface.\\We believe that, this discussion will provide a comprehensive understanding of the analytical behavior of the GH shift and its limitation. Contrary to the former researches, the most prominent features of this model are to clarify how and where the axial incident ray returns to the incident medium for angles equal to and greater than the critical angle and that the incident ray, as it should be, does not enter into the forbidden medium at all, since geometrical optics prohibits any penetration of light into an optically less dense medium in this case. We hope that, this work will encourage formulating new questions and experiments and help us to widen our horizons and lead to developing new physically relevant insights, interpreting data and designing further experiments.
\end{document}